**Stop Premature Obsolescence: LessTrash, Fewer Working Hours, Same Pay**

Tommaso Luzzati[1,3], J. Christopher Proctor[2], S. D'Alessandro[1]

[1]Dipartimento di Economia e Management - Università di Pisa, Italy; [2]Bocconi University - Institute for European Policymaking; [3]REMARC - Università di Pisa, Italy; tommaso.luzzati@unipi.it

**ABSTRACT**

About a century ago, Keynes predicted that, thanks to technological progress, today's working week would be reduced to 15 hours. In practice, however, large technological gains seem to have been offset by a substantial shortening of product lifespans. From a standard microeconomic perspective, this is inefficient as more labour and material resources are employed than necessary to deliver a given level of product-provided services. In this paper, we model policies aimed at reducing the current 'throwaway economy' as inducing a positive productivity shock in one economic sector and examine the conditions under which this shock's propagation to other sectors can allow for a reduction in working time, without reducing consumers' well-being, labour compensation, or profits. To this purpose, we first focus on a stylized three sector economy – using both a physically extended input-output example and a system-dynamic stock-flow model.

## 1. INTRODUCTION

The allocation of scarce means among alternative uses, as Lionel Robbins stated[1], represents the most common definition of economics and is the guiding principle of mainstream microeconomics. In macroeconomics this focus on efficiency is largely lost, particularly in growth theory and its reliance on GDP as an indicator of progress regardless of its composition – as highlighted by Daly (1998) in his notion of 'uneconomic growth'. Indeed, the economy has evolved into a 'throwaway economy', characterized by increased material throughput and waste generation per unit of service. Can this help explain why Keynes's prediction that the working week would by now be reduced to 15 hours has not come true (e.g., Luzzati et al. 2022)? In other words, to what extent, and under which conditions, can policies that target reducing the material use required to provide the same service lead to a reduction in working time without a loss in wage compensation? This is the central research question around which this short paper would like to stimulate scientific and policy debate.

While the nexus between material resource efficiency and waste or pollution is immediately visible (and central in the sustainability literature), our collective mindset barely recognises that 'producing for the bin' does not only involve dysfunctional material losses but also reflects a broader inefficiency in the use of all resources, including capital and labour. This issue is compelling and underexplored; it is highly relevant because, once production costs are expressed relative to the services a good can provide over its life cycle, a short-lived product is generally not cheaper than the corresponding long-lived one. From this perspective, we analyze, through analytical reflection and a series of simulations, how the multiple feedback effects generated by policies aimed at material savings propagate across economic sectors.

[1] "Economics is the science which studies human behaviour as a relationship between ends and scarce means which have alternative uses." (Robbins, 1932, p. 15)

While premature obsolescence refers to durable goods, non-durable goods also exhibit considerable potential for material savings. Here we consider a policy concerning a non-durable good. The reason is that the analysis for non-durable goods allows to minimize the elements of the argument, which can therefore be exposed in a very simple way. Our argument, however, readily extends to durable goods and to policies aimed at increasing product lifespans.

The paper is organised as follows. Section 2 discusses our example and the analytical elements at play. Section 3 shows a stylized Physically Extended Input–Output table linking monetary flows to quantities of productive factors (e.g., Towa et al., 2021). Section 4 adopts a system dynamics approach to build a stock-flow model which includes the feedback effects highlighted in previous sections. Section 5 discusses and concludes.

## 2. THE ARGUMENT

Let us consider the introduction of a law that mandates doubling the concentration of liquid detergents. Clearly, one could think of other policies to get similar effects, such as introducing a unit tax on plastic bottles, or a law banning liquid detergents, which is feasible as solid alternatives do not pose any particular technological challenge (e.g. Banti and Luzzati, 2026). Our example is chosen for the sake of simplicity as it involves a very simple analysis.

The first analytical step is to observe that such a policy can be modelled as a positive sector-specific technological shock that increases resource efficiency. To see this, note that the policy would not entail significant technical challenges or production costs as dilution levels are largely a matter of market strategy, which is confirmed by the simultaneous availability of highly concentrated and highly diluted liquid detergents. At the same time, it would generate business gains, through reduced packaging, storage, and transportation costs. Because the same bottle would deliver twice the cleaning services, firms could either maintain bottle size and raise the price or reduce bottle size while keeping the price unchanged. Either way, the policy would reduce production and transportation inputs, and, as a side effect, reduce emissions and plastic use (and subsequent waste). As a second step, we highlight the main channels through which such a sectoral shock propagates through the economy.

1) Intersectoral demand for the plastic packaging and transportation sectors would decline.

2) This would both reduce the intersectoral exchanges and put downward pressure on prices, turnover, and profits in the negatively affected sectors. Labour demand would decrease, generating a generalised downward pressure on hourly wages, the magnitude of which depends on the relative importance of the affected sectors and the size of the shock.

3) In the detergents industry, operating profits would increase, making market entry attractive to new firms, supported by capital released from other sectors. As the price elasticity of demand for detergents is likely to be relatively low (people will still need to wash the same amount of clothing or dishes), competitive pressure from new entrants will translate primarily into lower prices and only modest increases in quantity.

4) As a result, households would realise cost savings, which could be reallocated to the consumption of other goods and services (generating rebound effects in waste and pollution) and/or to reductions in working time. At the same time, these gains may be partially offset by income losses resulting from reduced employment and, to some extent, by downward pressure on economy-wide wages linked with the reduction in employment.

5) Concerning the distribution of the surplus generated by the policy, due to the increases in labour productivity significant hourly wage increases should accrue to workers in the detergents sector, thereby almost leaving the total labour compensation in this sector unchanged despite the reduction in working hours. In half the time, factory workers would bottle the same quantity of output, and truck drivers would deliver the same amount, implying a doubling of labour productivity. At the same time, workers would be attracted from other sectors, which would prevent substantial wage increases in the detergent sector and allow profits to rise further in the detergent sector, which in turn will feed back into capital adjustments across sectors, as highlighted in channel 3.

Which actual sectoral composition changes will occur and how the surpluses will be distributed will depend on the possibility for productive factors to move from one sector to another and on the relative bargaining power of the different agents involved. Accordingly, while the compensation of all productive factors might increase, the actual distribution of the generated value surplus will differ significantly depending on the forces at play and the institutional arrangements. In the following sections we will examine to what extent scenarios can feasibly exhibit unchanged labour compensation and profits alongside reduced working time and lower waste.

## 3. A SIMPLE PHYSICALL EXTENDED INPUT-OUTPUT REPRESENTATION

To illustrate the practical relevance of our argument, we start from a Physical Extended Input–Output tables linking monetary flows to quantities of productive factors, in a similar way to the Environmentally Extended Input Output tables (e.g., Towa et al., 2021). Our table covers only three sectors and is fictitious, although using plausible values for Italy. To reflect the policy's effects, we then adjust technical coefficients and factor requirements and derive alternative tables capturing economy-wide adjustments under different capital-labour bargaining power balances.

The initial situation is roughly calibrated on Italy's economic structure in the years 2021-22 but normalized for a GDP of 2000 billion euros (2130 billion euros was the 2023 figure). This is done merely because a round GDP figure helps reading the different tables[2]. The initial monetary INPUT-OUTPUT table is as follows in Table 1.

**Table 1. Initial state of the system**

| | I | II | III | | Consum. | Invest. | eXports | *Final demand* | *Total* |
|---|---|---|---|---|---|---|---|---|---|
| **I** | € 5 | € 23 | € 19 | | € 33 | € 2 | € 18 | *€ 53* | *€ 100* |
| **II** | € 17 | € 732 | € 300 | | € 357 | € 300 | € 386 | *€ 1 043* | *€ 2 092* |
| **III** | € 21 | € 282 | € 235 | | € 1 126 | € 170 | € 281 | *€ 1 577* | *€ 2 115* |
| **Labour comp** | € 17 | € 312 | € 639 | *→ € 969* | *€ 1 517* | *€ 472* | *€ 684* | *€ 2 673* | |
| **Other VAs** | € 19 | € 397 | € 615 | *→ € 1 031* | | | *net X € 11* | | |
| **Imports (M)** | € 21 | € 345 | € 307 | - - - - - - | - - - - - - | - - - - - - | - - - - - -▶ | *- € 673* | |
| | | | | ***€ 2 000*** | ◀- - - - | **GDP** | - - - -▶ | ***€ 2 000*** | |
| *Tot Dom Prod* | *€ 79* | *€ 1 747* | *€ 1 808* | | | | | | |
| *Total* | *€ 100* | *€ 2 092* | *€ 2 115* | | | | | | |

In our exercise, this monetary table is thought as resulting from physical values and prices, as shown in Table 2.

The policy that doubles the efficiency of sector II means that the sector needs only half of the initial physical quantities (Table 2). This implies a new Leontieff's inverse matrix that allows to take into account the changes in the intersectoral exchanges. For the initial demand level, this leads to the new situation shown in Table 3.

**Table 2. Initial state of the system with prices and quantities explicit**

| | I | II | III | Consum | Invest | Exports |
|---|---|---|---|---|---|---|
| | *Q P* | *Q P* | *Q P* | *Q P* | *Q P* | *Q P* |
| **I** | 1.4 €3.5 | 6.6 €3.5 | 5.5 €3.5 | 9.5 €3.5 | 0.6 €3.5 | 5.0 €3.5 |
| **II** | 2.0 €8.5 | 86.2 €8.5 | 35.3 €8.5 | 42.0 €8.5 | 35.3 €8.5 | 45.5 €8.5 |
| **III** | 2.4 €8.5 | 33.2 €8.5 | 27.6 €8.5 | 132.5 €8.5 | 20.0 €8.5 | 33.0 €8.5 |
| **Labour comp** | 5.0 €3.5 | 24.0 €13.0 | 71.0 €9.0 | Work hrs | **100.0** | |
| **Other VAs** | 1.3 €15.0 | 26.5 €15.0 | 41.0 €15.0 | Other VAs | **68.8** | |
| **Imports (M)** | 6.02 €3.5 | 40.6 €8.5 | 36,1 €8,5 | | | |

**Table 3. Productivity in the secondary sector (II) doubles and sectoral exchanges change according to the new technical coefficients and resulting Leontieff's inverse**

| | I | II | III | | C | I | X | *Final demand* | *Total* |
|---|---|---|---|---|---|---|---|---|---|
| **I** | € 4 | € 9 | € 17 | | € 33 | € 2 | € 18 | *€ 53* | *€ 89* |
| **II** | € 14 | € 282 | € 271 | | € 357 | € 300 | € 386 | *€ 1 043* | *€ 1 726* |
| **III** | € 17 | € 109 | € 213 | | € 1 126 | € 170 | € 281 | *€ 1 577* | *€ 1 973* |
| **Labour comp** | € 15 | € 120 | € 579 | *→ € 713* | *€ 1 517* | *€ 472* | *€ 684* | *€ 2 673* | |
| **Other VAs** | € 16 | € 958 | € 557 | *→ € 1 531* | | | *Net X € 256* | | |

[2] In some cases, rounding causes very small inconsistencies in the figures (sums or products).

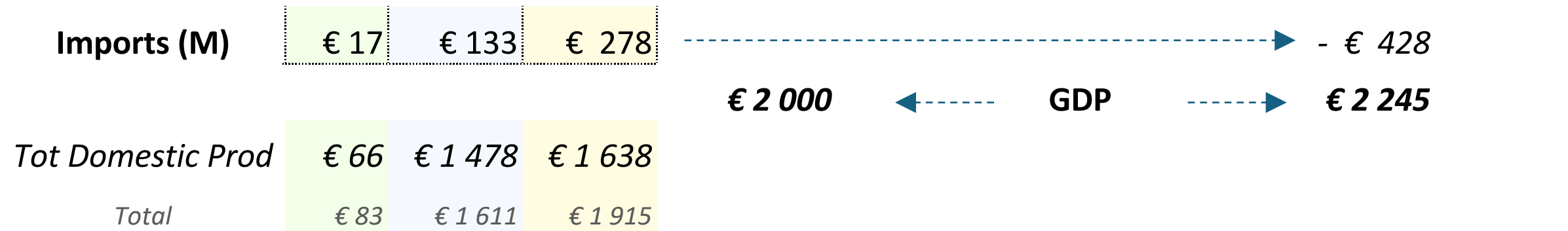

| | I | II | III | | | | |
|---|---|---|---|---|---|---|---|
| **Imports (M)** | € 17 | € 133 | € 278 | ------------------------------------------► | | | - € 428 |
| | | | | **€ 2 000** | ◄------ | GDP | ------► **€ 2 245** |
| *Tot Domestic Prod* | *€ 66* | *€ 1 478* | *€ 1 638* | | | | |
| *Total* | *€ 83* | *€ 1 611* | *€ 1 915* | | | | |

As sector II experienced a positive productivity shock, and assuming that labour compensation does not change, the rents and profits of the other factors (VA over their quantity) experienced a sharp increase as shown in Table 4, which also shows the decrease in working hours across the sectors.

**Table 4. New Rents and working hours, after the shock, with unchanged wages and prices**

| Rents | Wages | Final P | Working hrs | w.r.t. initial hrs |
|---|---|---|---|---|
| 15.0 € | 3.5 € | 3.5 € | 4.2 | 83% |
| 94.1 € | 13.0 € | 8.5 € | 9.2 | 38% |
| 15.0 € | 9.0 € | 8.5 € | 64.3 | 91% |
| | | | 77.7 | 77.7% |

This differential will induce a process of mobility of all factors that will end when the rents across the three sectors are equalised, and wages will increase, keeping however their relative differences across sectors. The final endpoint clearly depends, ceteris paribus, on the different bargaining power of the workers as compared to rentiers or capitalists. Among the many possible outcomes, we show below two different scenarios in which wages increase respectively by 15% and by about 69,7%. The latter figure The latter figure represents, in our example, the percentage required to maintain the pre-shock distribution of GDP between labour and other factors. For the sake of simplicity, we calibrate changes in final prices - given a vector of price elasticities of demand - so that GPD remains unchanged at 2000 billion euros. The first example is shown in Table 5 and the second in Table 6.

**Table 5 Scenario A: Changes in prices to equalize rents between sectors and wage increases by 15%**

| | I | II | III | | C | I | X | *Final demand* | *Total* |
|---|---|---|---|---|---|---|---|---|---|
| **I** | € 5 | € 12 | € 19 | | € 42 | € 3 | € 24 | *€ 69* | *€ 105* |
| **II** | € 7 | € 148 | € 115 | | € 203 | € 163 | € 210 | *€ 575* | *€ 845* |
| **III** | € 23 | € 153 | € 242 | | € 1 107 | € 248 | € 409 | *€ 1 764* | *€ 2 183* |
| **Labour comp** | € 15 | € 134 | € 520 | *→ € 669* | *€ 1 352* | *€ 413* | *€ 642* | *€ 2 408* | |
| **Other VAs** | € 32 | € 329 | € 970 | *→ € 1 331* | | | *Net X € 234* | | |
| **Imports (M)** | € 22 | € 70 | € 316 | ------------------------------------------► | | | | *- € 408* | |
| | | | | **€ 2 000** ◄------ | GDP | ------► | | **€ 2 000** | |
| *Tot Dom. Prod* | *€ 83* | *€ 775* | *€ 1 866* | | | | | | |
| *Total* | *€ 105* | *€ 845* | *€ 2 183* | | | | | | |

| Other VAs | Wages | Final P | Initial P | Working hrs | w.r.t. initial hrs |
|---|---|---|---|---|---|
| 33.4 € | 4.0 € | 4.8 € | €3.5 | 3.8 | 76% |
| 33.4 € | 15.0 € | 4.6 € | €8.5 | 8.9 | 37% |
| 33.4 € | 10.4 € | 12.4 € | €8.5 | 50.3 | 69% |

| | | 63.0 | 63% |
|---|---|---|---|

**Table 6 Scenario B: Changes in prices to equalize rents across sectors and wage increases by 69,7%**

| | I | II | III | | C | I | X | *Final demand* | *Total* |
|---|---|---|---|---|---|---|---|---|---|
| **I** | € 5 | € 12 | € 18 | | € 43 | € 3 | € 24 | *€ 70* | *€ 106* |
| **II** | € 7 | € 147 | € 112 | | € 202 | € 162 | € 209 | *€ 574* | *€ 840* |
| **III** | € 23 | € 157 | € 243 | | € 1 090 | € 254 | € 420 | *€ 1 765* | *€ 2 188* |
| **Labour comp** | € 22 | € 196 | € 750 | *→ € 969* | *€ 1 335* | *€ 420* | *€ 653* | *€ 2 408* | |
| **Other VAs** | € 25 | € 259 | € 748 | *→ € 1 031* | | *Net X* | *€ 245* | | |
| **Imports (M)** | € 22 | € 69 | € 317 | | | | | *- € 408* | |
| | | | | ***€ 2 000*** | | **GDP** | | ***€ 2 000*** | |
| *Tot Dom. Prod* | *€ 83* | *€ 775* | *€ 1 866* | | | | | | |
| *Total* | *€ 106* | *€ 840* | *€ 2 188* | | | | | | |

| Other VAs | Wages | Final P | Initial prices | Working hrs | w.r.t. initial hrs |
|---|---|---|---|---|---|
| 26.4 € | 5.9 € | 4.9 € | €3.5 | 3.8 | 76% |
| 26.4 € | 22.1 € | 4.6 € | €8.5 | 8.9 | 37% |
| 26.4 € | 15.3 € | 12.7 € | €8.5 | 49.1 | 71% |
| | | | | 61.8 | 61.8% |

## 4. A DYNAMIC MODEL

The second step is to develop a system dynamic approach by building a stylized stock-flow model which includes the feedback effects described in Section 2, that is, accounting for rent equalization across sectors, wage setting based on productivity changes, and demand responses to changes in relative prices and to changes in aggregate income. This gives a richer representation of the stylistic Input Output analysis described in Section 3 and allows for the development of some hypothetical scenario analysis.

### 4.1 Framework

The causal loop diagram in Figure 1 gives an overview of the main positive and negative relationships represented in the model, which will be described in this section. Starting with final demand, as the demand for a given sector increases, the output of that sector will also increase. The output for a sector will also increase when more inputs are needed to create the same level of output, or as in our example, the reverse.

As levels of output increase, more labor is required. Labor productivity is considered as the amount of output per hour worked, and as such it will go up when either output increases or working hours decrease. In our example, these two channels will partially offset each other, as output and working hours both initially decrease due to the productivity shock. Labor productivity is also impacted indirectly by changes to the labor intensity of production, which directly impacts labor use. Economy-wide wages increase along with increases in labor productivity within each sector. Labor compensation is the combination of hours worked and wages and moves positively with both. Profits are considered to be everything leftover after capitalists have paid workers (labor compensation) and purchased the inputs required for production. As such profits increase as required inputs or labor compensation decrease. Profits also increase as outputs increase, or as prices increase. Rents (the rate of return on capital in that sector) increase as profits increase, or as the capital intensity of production decreases. As rents in a sector increase, capital from other sectors will flow in to take advantage of the high rates of return. After some time (as indicated by the dashed "delay" mark), the increased levels of investment in a sector will put downward pressure on prices for that sector's output. This reduction in prices will likewise reduce the profits in the sector.

This loop between profits, rents, investment and prices is the first balancing loop in the system (labeled B1) and represents the tendency of rents to equalize between sectors, with the equalization process occurring through an eventual reduction in prices for sectors with above-average rents. This loop was represented statically in Tables 4 and 5 above with the changes in prices to equalize rents, given a particular wage level.

The remaining loops determine the development of final demand. First, there are first reinforcing loops which represent how increases to demand feed increases to national income, with R1 indicating that an increase in profits will allow for an increase in economy-wide investment, and R2 and R3 showing how increases in labor compensation increase consumption, either through more hours worked (R2) and higher wages (R3).

Finally, as relative prices change between sectors (in response to changes in rents), the relative demand between sectors will also fluctuate, with a decrease in prices associated with an increase in demand. As demand increases, output, profitability and rents in a sector all increase, leading to capital inflows and reduced prices, creating a final reinforcing loop (R4).[3]

The key exogenous variables in this system are the inputs needed for production and the labor and capital intensities of production, all of which we conceptualize as being reduced by a policy-mandated increase in concentration, as fewer physical inputs, working hours, and machines are needed to create each unit of product.

[3] In our specific example of laundry detergents, we imagine this reinforcing loop via changes in relative prices would be rather weak, as the elasticity of demand for laundry detergents is quite low (people will only demand so much soap, even if it becomes very cheap). This may not be the case for other products. In particular, for durable goods, price effects could be very important, motivating the inclusion on this feedback loop in our framework.

**Figure 1 A casual loop diagram of the dynamic model.**

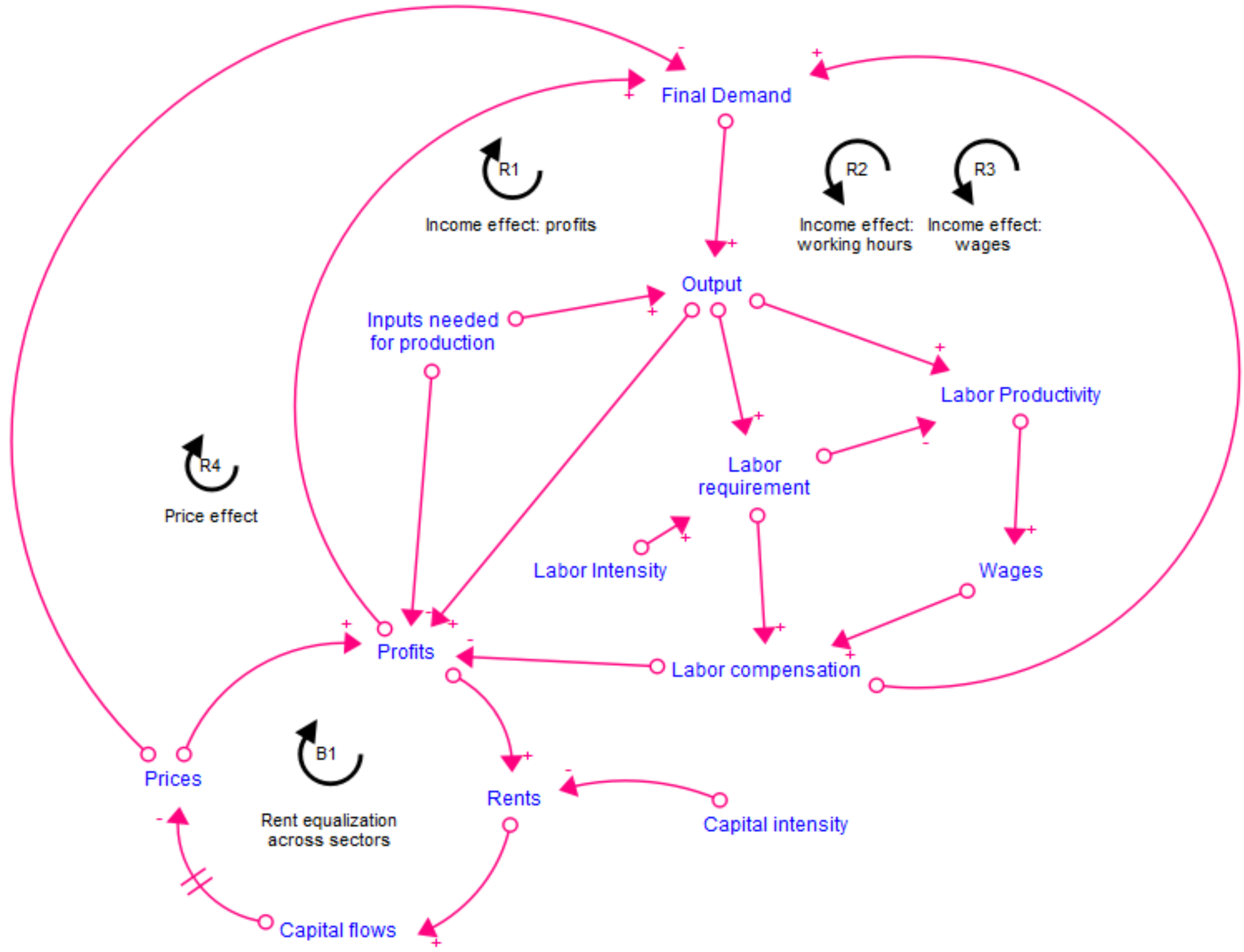

## 4.2. Model

The modelization of this CLD replicates the simple input-output framework described above for the calculation of output (in quantities), with the technical coefficients of production (A) transformed into the Leontief Inverse $(I-A)^{-1}$ and then multiplied by the vector of final demand. Labor, capital and imports required are calculated via production intensities, which are multiplied by output for each sector.

Profits (labeled as "other value added") are calculated by subtracting intermediate inputs, labor compensation and imports from gross output. Rates of rent are calculated by dividing profits of a sector by its required capital, which is in turn calculated by multiplying the capital intensity of each sector by its output.

The price for the goods in each sector adjusts gradually in order to equalize the difference between that sectors' rent and the average rent. This is done by increasing or decreasing the sector's price by a given 'rate of rent equalization' in each year that its rent is over or under the average rate of rent by a certain threshold.[4]

[4] The threshold for rent equalization, along with the rate of rent equalization determine the path of convergence of prices, profits, and rents after a productivity shock. In the current initialization, these parameters are responsible for wave-like fluctuations as each sector approaches the average rent rate, but is not able to fully converge to precisely it. A lower rate of rent equalization or a higher threshold would allow for convergence but would take much longer to do so.

Wages adapt directly to changes in labor productivity (output of value per hour) based on the bargaining power of workers relative to capital (an arbitrary value between 0 and 1), with the bargaining power multiplied by the productivity change in percentage to determine the percentage change in wages.[5]

Final demand in quantities adapts to both changes in income and changes in relative prices. For the income effects, changes to labor compensation impact final consumption, while changes to profits impact investments. The change in consumption is modeled by multiplying the log of the annual change in total labor compensation by the level of consumption. Change in investment is similarly the product of the log of the annual change investment by the level of investment.   Exports are not subject to income effects.

Price effects are modeled by multiplying the negative log of changes in prices for each sector by final demand for a sector (consumption, investment and exports), and a price elasticity for each sector (set between 0 and 1). This is intended to capture an approximate movement of demand between sectors as relative prices change, with an understanding that the salience of these effects will be quite low, especially in the primary and secondary sector.[6] The initial values for the quantities of final demand, sectoral prices, intensities (labor, capital, and import) and technical structure (A matrix) are derived from example in Table 1.

### 4.3 Scenarios

The policy shock in the model is designed as a 50% reduction to the column of the technical coefficients (A) representing the inputs used by the secondary sector, and the labor, capital and import intensities of the secondary sector. As in Section 3, this is an extreme example used to clarify the dynamics associated with a policy mandating the increased concentration of laundry detergents.

First, we will present a scenario in which only half of the productivity gains from the shock are translated into wage increases, corresponding roughly to Scenario A in Section 3. Then we will show a comparison with a scenario in which productivity gains are fully translated into wage increases, representing Scenario B above. The first dynamic to observe is the sharp reduction in output across all sectors due to the reduction in input use in the secondary sector (Figure 2). In the secondary sector, the bulk of the reduction occurs in the first year after the shock, due to the very large amount of input flows within the secondary sector itself (see Table 1 where secondary-secondary is by far the largest source of input use). In the primary and tertiary sectors, output falls more gradually, due to indirect effects.

---

[5] Like in the input-output example above, wages between sectors tend to converge to maintain the same relative ratio as in the initial state. This comes from changes in the labor productivity of each sector, not from an explicit modeling of wage convergence to the starting ratio.

[6] Following the example from Section 3, the price elasticities for the primary, secondary and tertiary sectors are set a .2, .1, and .7 respectively.

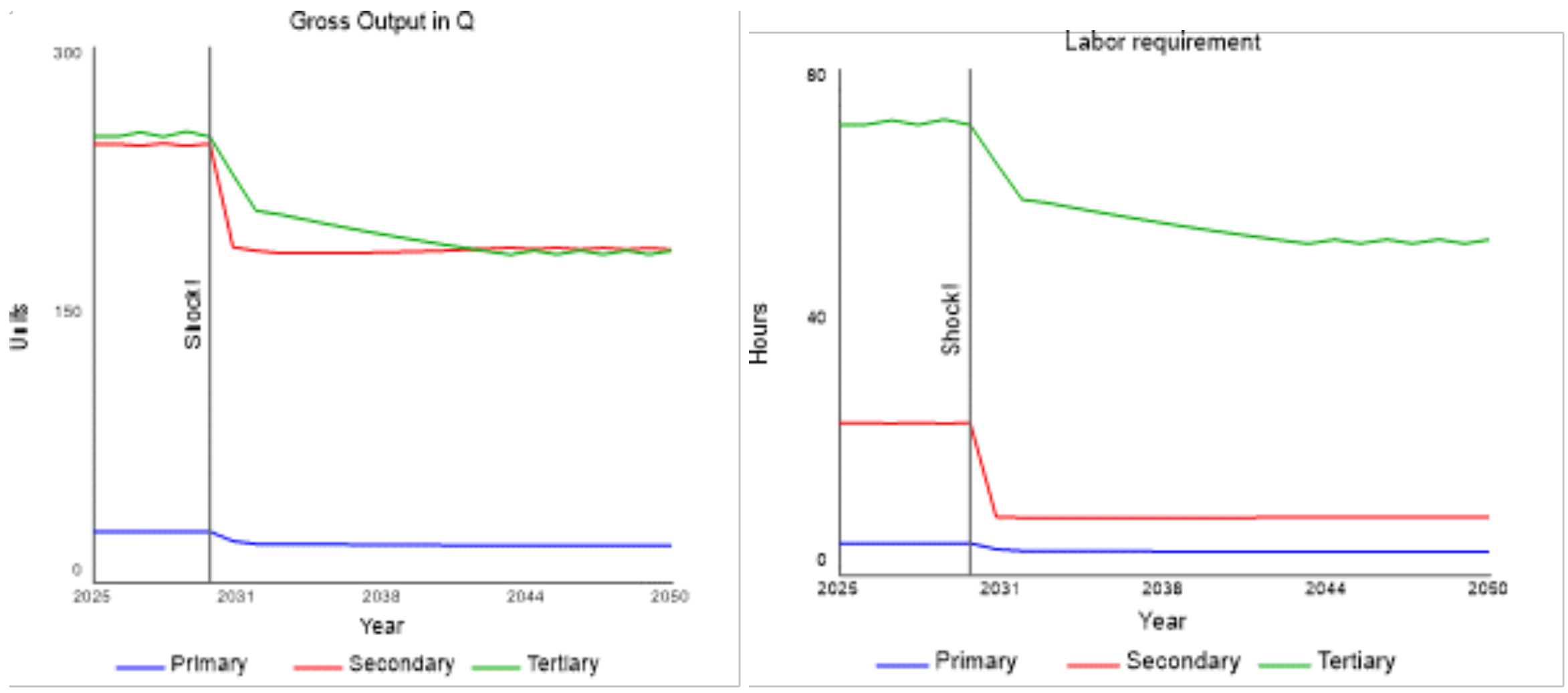

**Figure 2. Output falls as fewer inputs are needed**

**Figure 3. Working hours fall, especially in the secondary sector**

Working hours also fall substantially after the policy shock (Figure 3). For the primary and tertiary sectors, the decline in labor use directly follows the decline in output – fewer workers are needed as less is being produced. In the secondary sector, the effect from the reduction of output is combined with an additional effect from the increase in labour productivity, as the newly concentrated detergent requires half as much labour to fulfill the same amount of demand.

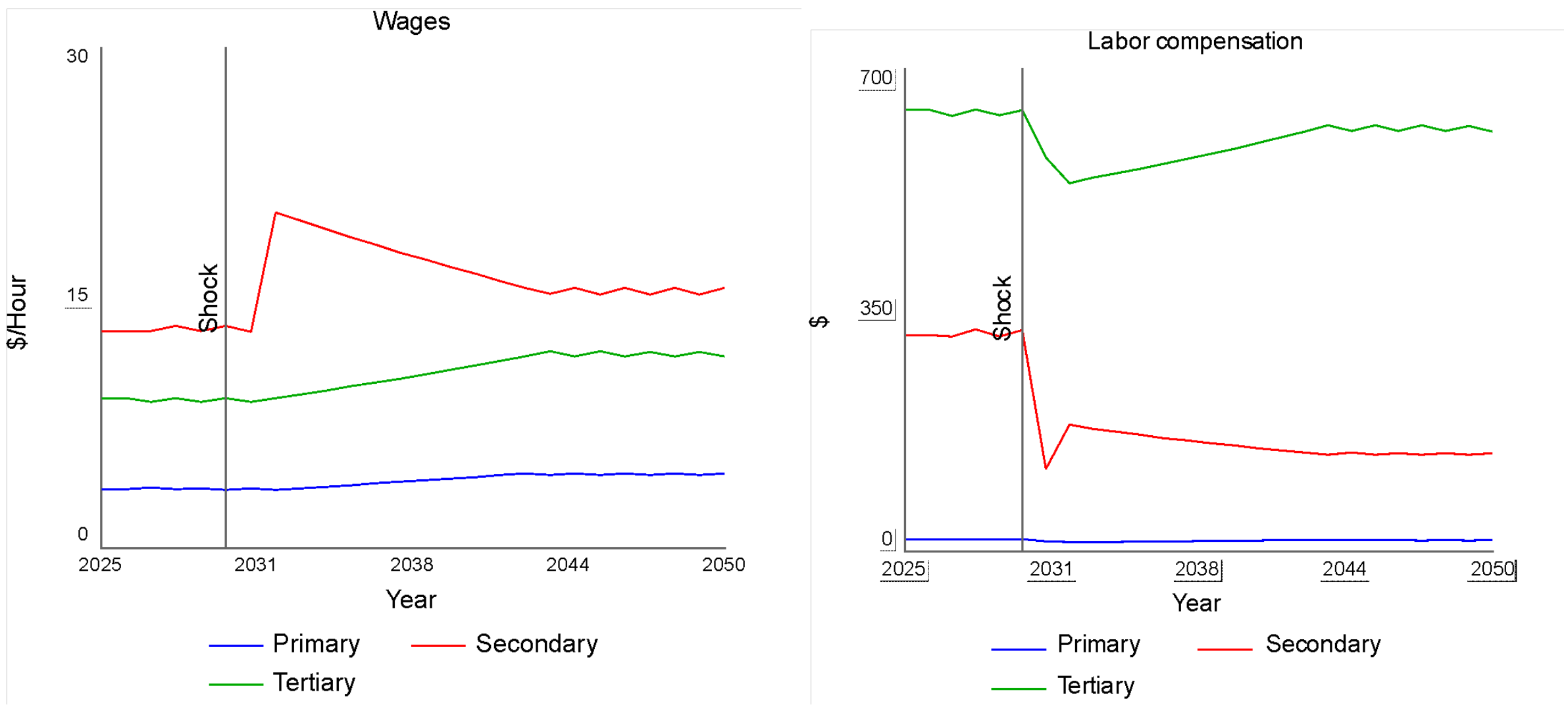

**Figure 4. Wages converge upwards, maintaining a similar ratio**

**Figure 5. Labor compensation falls in all sectors but recovers in Tertiary**

In response to the increase in labor productivity in the secondary sector, wages increase sharply in that sector (Figure 4). This is followed by a gradual convergence due to subsequent changes in prices (Figure 8) as wages increase in the primary and tertiary sectors in line with price increases in those sectors and fall in the secondary sector as prices in that sector decrease. The wage structure eventually converges at approximately the same wage-ratio between the three sectors as in the initial state, but at a higher level. Total labor compensation (Figure 5) falls in all sectors and then partially recovers in the primary and tertiary sectors.

Profits ("other value added") spike in the secondary sector with the reduction in production costs (inputs, labor, capital, imports) (Figure 6). Secondary sector profits then gradually fall due to falling prices. In parallel, primary and tertiary sector profits increase along with raising prices. Primary sector profits, while small overall, roughly double after the shock.

Rents in the secondary sector spike (Figure 7), as both profits increase and capital intensity decrease (similarly to labor, half as many machines are needed to fill the same demand for detergent). Rent rates are then forced to converge by the model, with a convergence rate set such that full convergence takes roughly 10 years, giving time to more easily see the dynamics before, during and after the convergence.

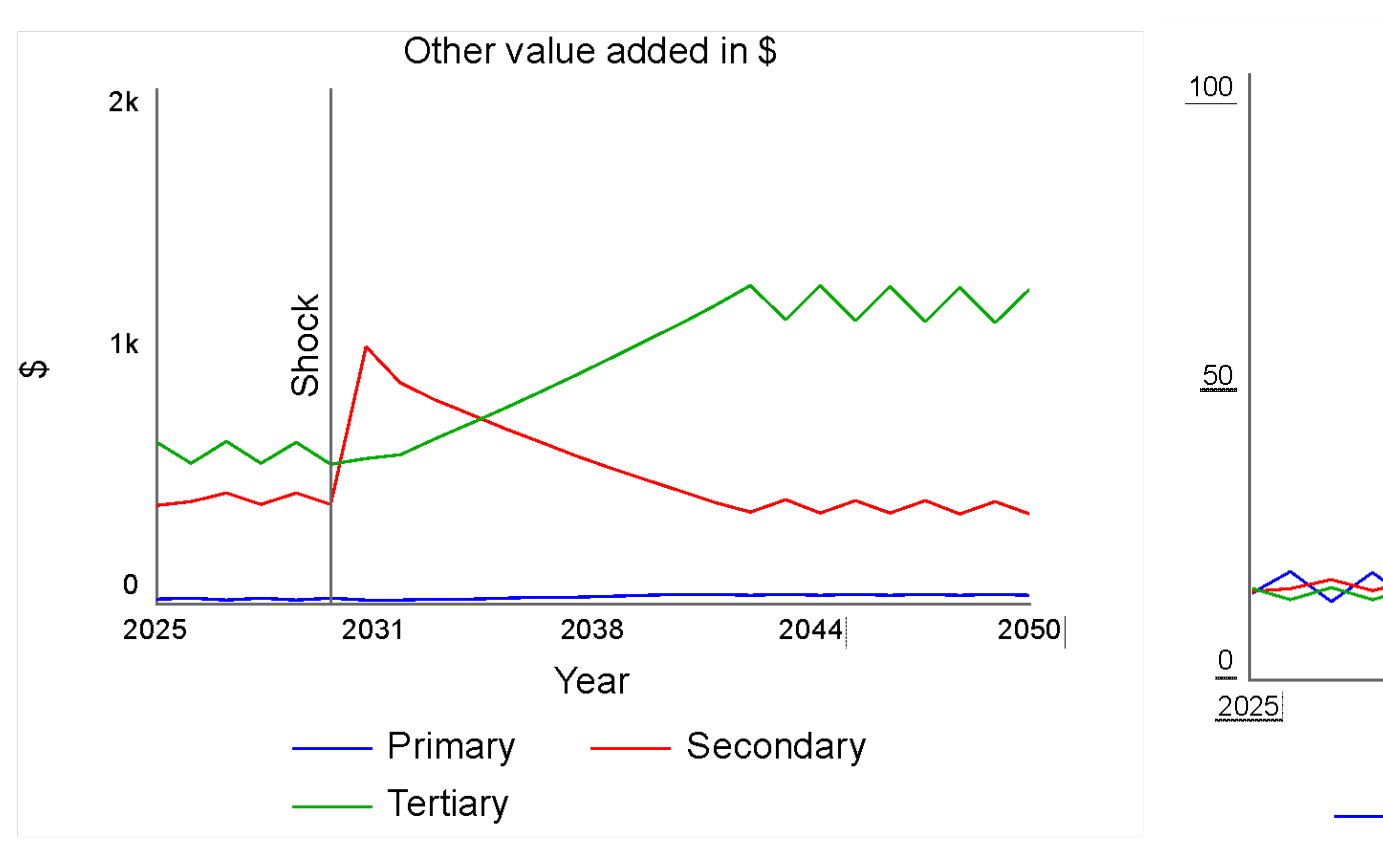

**Figure 6. Profits increase first in the secondary sector, then in the tertiary**

**Figure 7. Rents spike in the secondary sector and then converge**

Rent equalization is achieved through changes to prices (Figure 8) which increase in the primary and tertiary sectors and decrease in the secondary sector before stabilizing once rents have equalized. Real final demand (in quantities) falls overall (Figure 9), along with the fall in labor compensation, but increases in the secondary sector due to the sharp fall in relative prices.

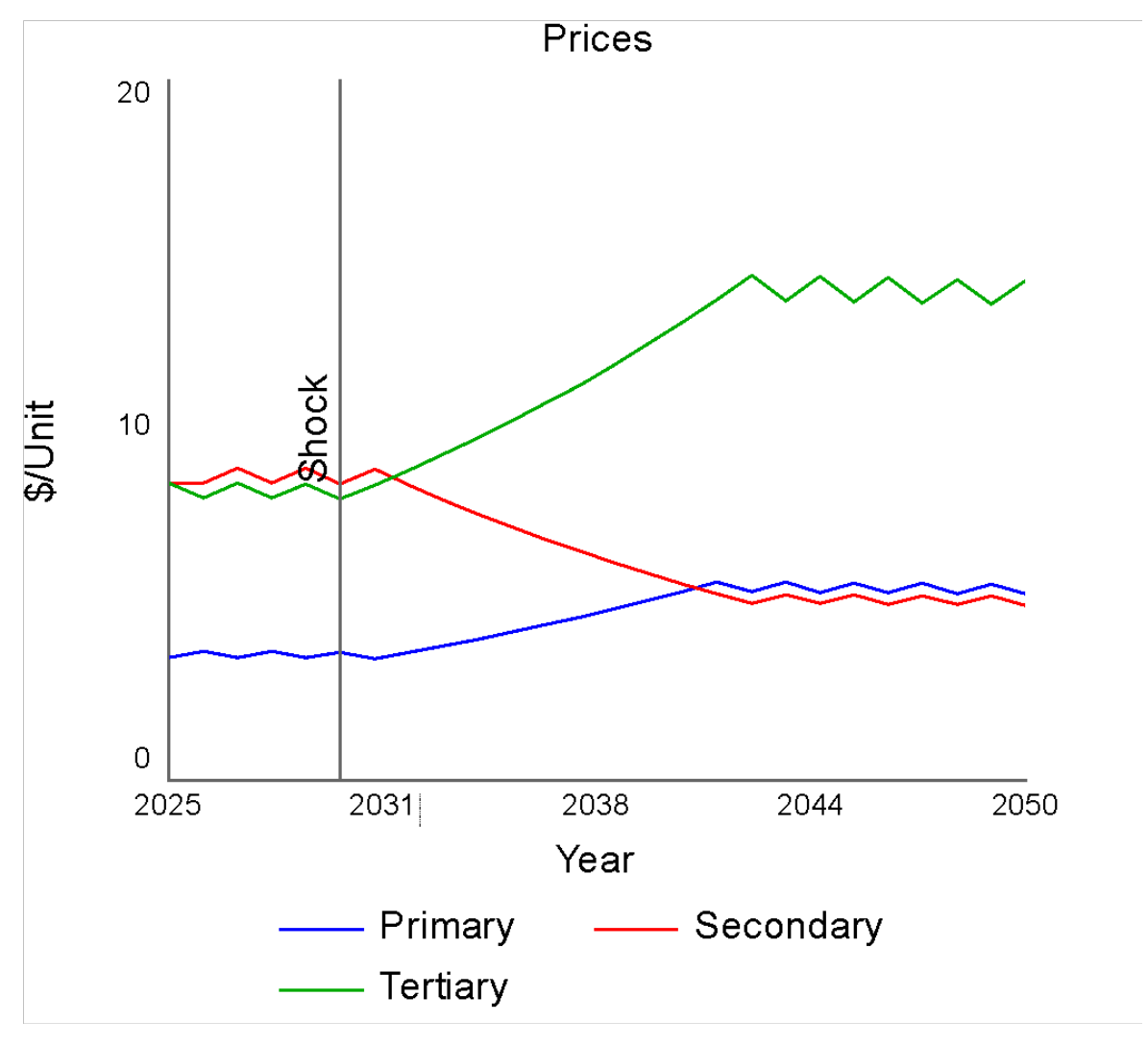

**Figure 8 Prices adjust to equalize rents across sectors**

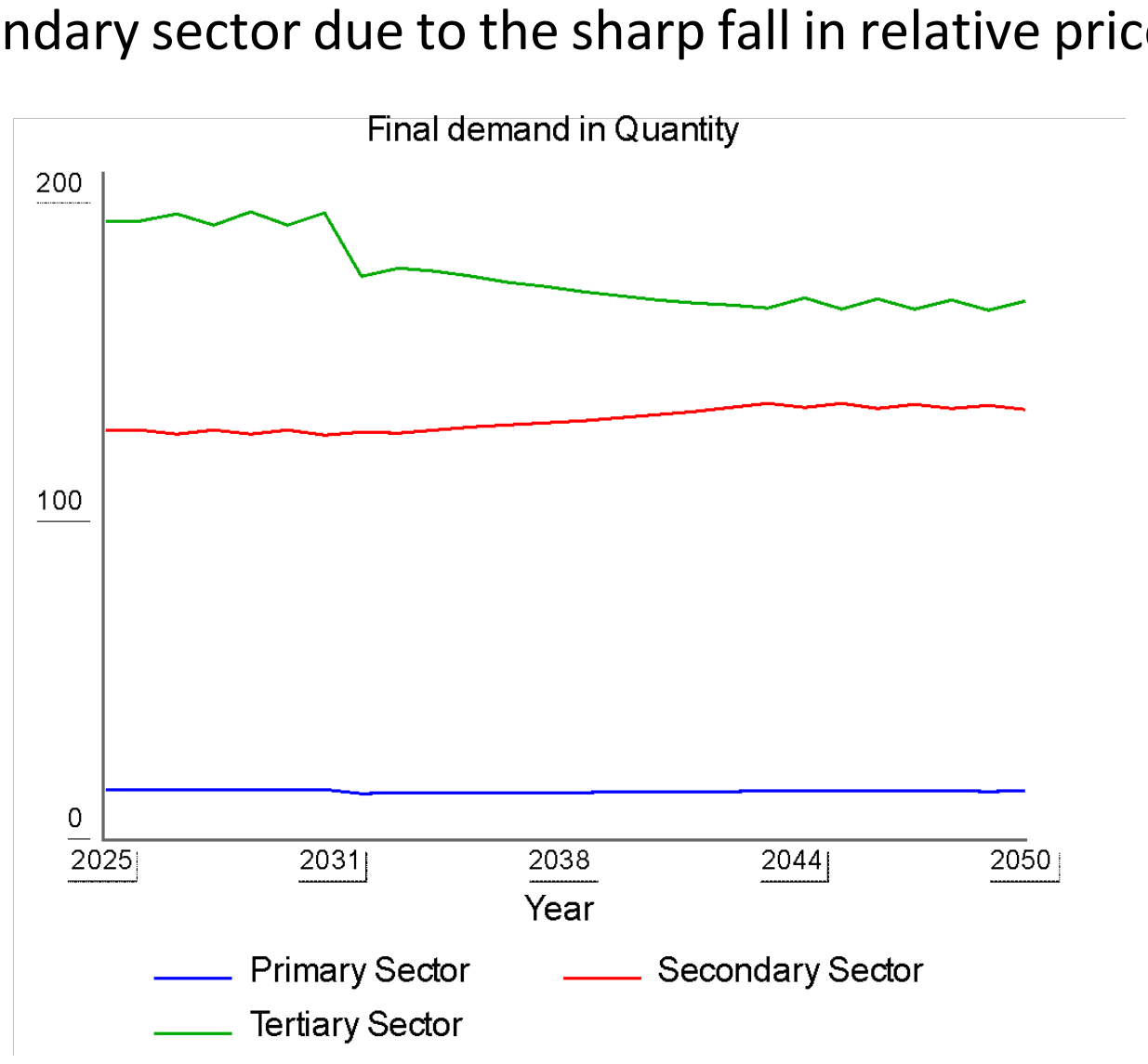

**Figure 9 Final demand declines with falling labour compensation and adjusts with prices between sectors**

In a scenario where workers have a stronger bargaining power to fully increase wages along with productivity increases (the 100% scenario in Figure 10), wages in the secondary sector increase much higher after the initial shock, before falling towards the same level as in the previous scenario. Wages in the primary and tertiary sectors however increase permanently. In this scenario, the starting ratio of relative wages between the sectors is not maintained, with tertiary sector wages nearly catching the secondary sector. The higher wages across all sectors lead to an increase in overall labor compensation and a decline in profits (Figure 11), with total labor compensation reaching roughly the same level as in the initial state. This is without a change in the dynamics of working hours, which remain essentially the same in both scenarios (Figure 12), with a 35% decline in Scenario A and a 30% decline in Scenario B.

The slightly higher level of working hours in the scenario with higher worker bargaining power is due largely to the positive impact of higher wages on final demand, which increases more with increases in labor compensation than it declines with decreases in profits. This can be seen in the higher level of GDP in Scenario B in Figure 13. In both scenarios, the upward trend in GDP is largely driven by the reduction of imports used in production, which are not offset by a reduction in exports.

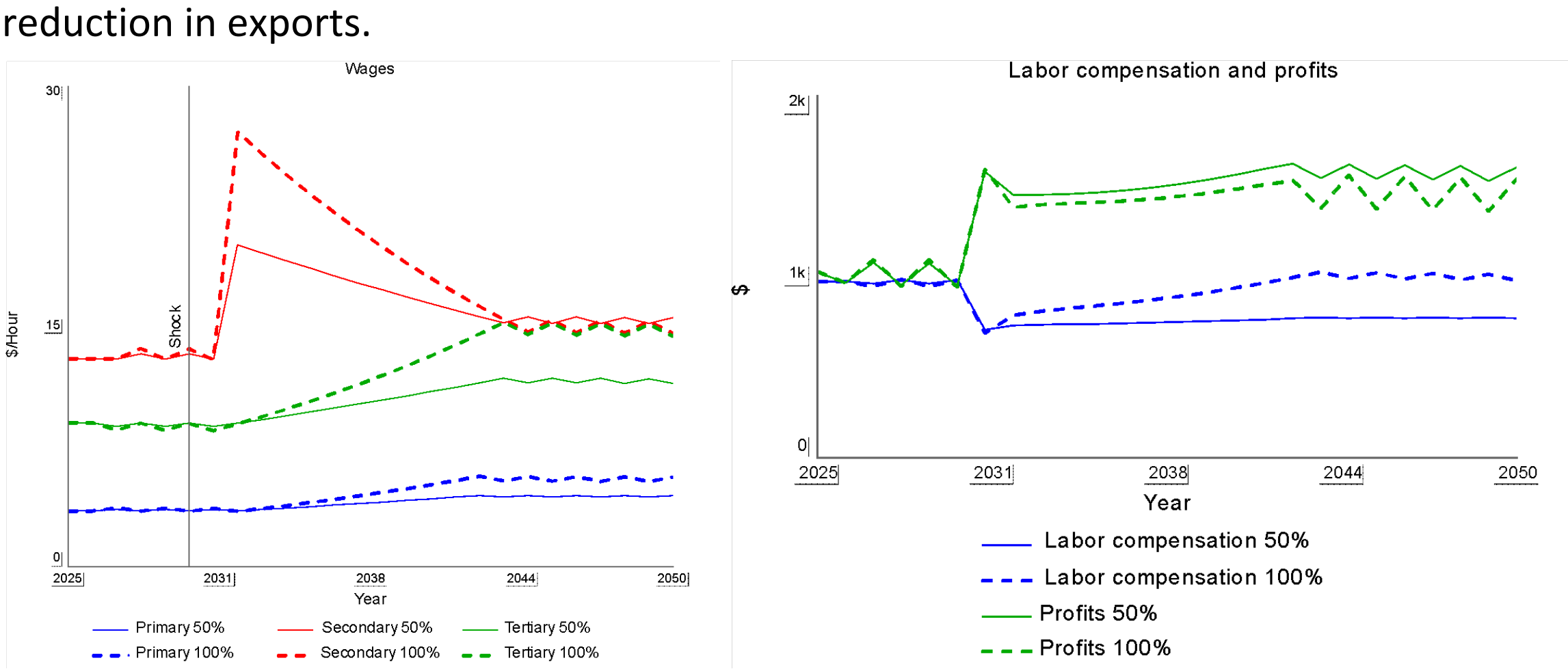

**Figure 10 Wages are higher in the secondary sector when 100% of productivity gains go to increased wages**

**Figure 11 Labor compensation is higher (and profits lower) with higher worker bargaining power**

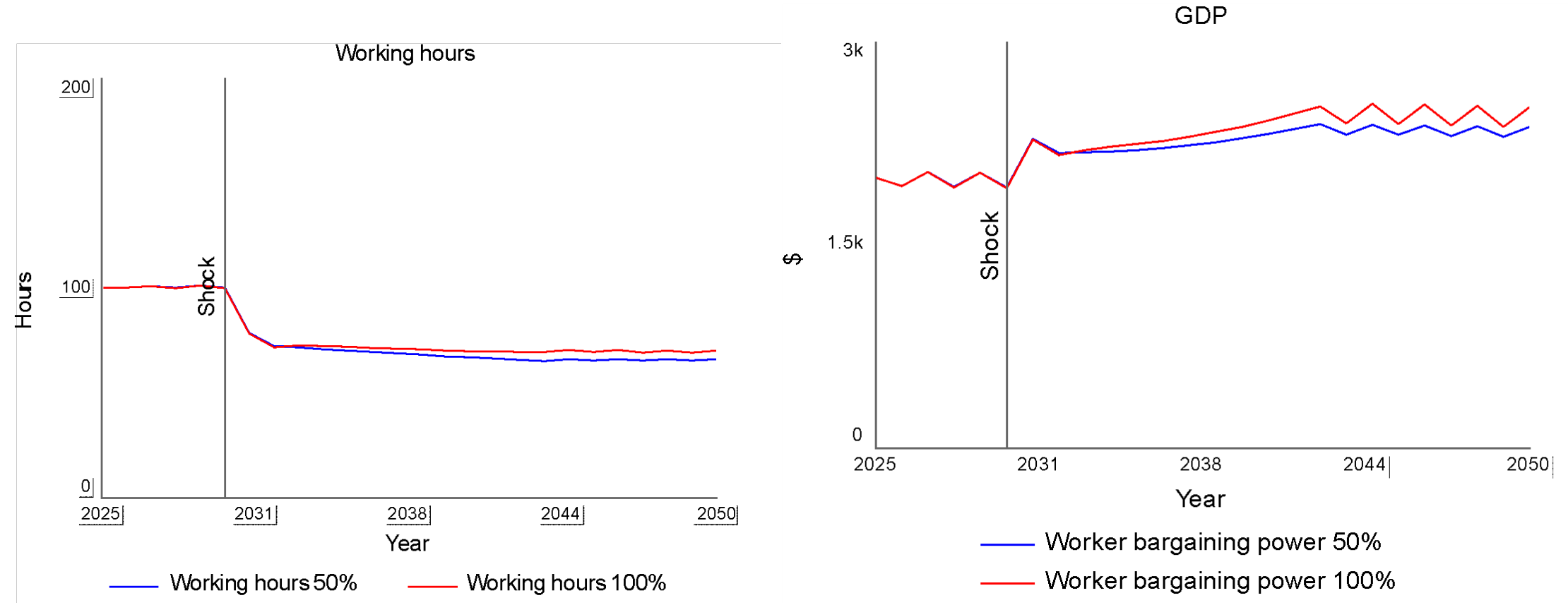

**Figure 12 Working hours decline similarly in both scenarios**

**Figure 13 GDP is higher as the income effect for consumption overpowers the income effect for investments**

## 5. SCENARIOS IN THE EUROGREEN MODEL

The natural continuation of this research is to develop its arguments within a detailed real-economy model, such as EUROGREEN (D'Alessandro et al., 2025). Within this model, it is possible to introduce a sector-specific productivity shock (in the 'petro-chemical' and 'detergents & packaging' sectors) and to trace its propagation through the input–output structure. Developing our argument within a real-economy model would allow us to derive alternative scenarios reflecting different productivity gains and working-time reduction (WTR) intensities, and to track prices, employment, total labour compensation, profits, inequality, waste and material throughput, emissions, and the public budget. It would also allow us to assess the extent of rebound effects. In our hypothetical case, this appears limited, as the improvement in efficiency mainly translates into reductions in working time, without increases in income and with changes in the price vector that do not allow income savings to be directed toward higher consumption and the further stimulation of economic activity. Clearly, this result depends on the assumption that we did, in particular the price elasticities of the goods and services.

This section will be completed soon.

## 6. CONCLUSION

As stated in the introduction, our models are highly stylized, as the focus is on mechanisms and qualitative insights. For this reason, we use the example of concentrating diluted liquid detergents rather than analysing the extension of product lifetimes in more material- and water-intensive goods (e.g., household appliances). Nonetheless, liquid detergents still entail non-negligible impacts across all stages of the life cycle (e.g., De Moura et al., 2023). That said, focusing on a consumable, allows us to avoid introducing into the analysis of dynamic elements associated with durable goods. In addition, producing less diluted detergents would not require significant technological changes or higher costs; on the contrary, the opposite is true. Most importantly, and in contrast to the existing literature, our focus here is not on material resources or environmental impacts, but rather on the fact that the "throwaway economy" also entails substantial labour waste. Addressing this inefficiency would allow for a significant reduction in working time without reducing labour compensation or the services provided by the economy.

Clearly, our analysis does not allow us to make quantitative predictions as we applied Occam's razor to an extreme by basing our analysis on a fictitious three-sector economy and assuming perfect mobility of factors. This was made to illustrate the argument in the simplest possible way.

From a policy perspective, an implication of our analysis is that he persistence of material inefficiencies, despite their apparent technical and cost-saving potential, reflects their role as equilibrium outcomes shaped by competitive pressures rather than isolated firm-level decisions. As in many cases of premature obsolescence, multiple barriers persist (Gulati et al., 2025), including insufficient incentives for firms to unilaterally depart from prevailing market practices when such departures may erode market share or profitability.

The policy implication is straightforward: efficiency-enhancing changes that reduce material throughput and extend product lifetime are unlikely to emerge endogenously at the scale required. Instead, they require coordinated policy interventions - such as standards on product concentration, durability, or reparability - that effectively redefine the competitive baseline across entire sectors.

The main contribution of our paper is to show that such interventions has be evaluated not only through the lens of environmental benefits, but also as macroeconomic interventions capable of reshaping the allocation of labour, income, and time. By inducing sector-specific productivity gains, material efficiency policies can generate economy-wide value surpluses that, under appropriate institutional conditions, can be redirected toward reductions in working time without compromising total labour compensation or firm profitability. The distribution of these gains is not automatic but depends on the bargaining power of workers including both wage setting powers and complementary institutional factors such as working-time regulations and social protection systems. Accordingly, effective policy design must combine material efficiency standards with measures that actively steer the use of the resulting surplus. In this sense, tackling the "throwaway economy" is not only an environmental imperative but also an opportunity to revisit long-standing expectations – such as Keynes's vision of shorter working weeks – and align technological progress with improvements in well-being and time sovereignty.

## ACKNOWLEDGMENTS

This research was developed as a part of the project "Well-being In a Dematerialized Economy (WIDE)" funded by the Italian Ministry of University and Research - PNRR M4 C211-CodeP2022S83EA-CUPI53D23006770001.
This manuscript is the result of an intense and equal collaboration among the authors, with Luzzati drafting section 1, 2, 3, 5 and Proctor section 4.
We gratefully acknowledge the valuable comments provided by the anonymous referees of the Ghent European Society for Ecological Economics Conference, to be held in June 2026.